\documentclass[prd,aps,floatfix,nofootinbib,11 pt]{revtex4}
\usepackage{amsmath,graphicx,color,epsfig}

\input{tcilatex}

\begin{document}

\title{Entanglement dynamics of a qubit-qutrit system in non-inertial frames}
\author{M. Ramzan\thanks{%
mramzan@phys.qau.edu.pk}}
\address{Department of Physics Quaid-i-Azam University \\
Islamabad 45320, Pakistan}
\date{\today}

\begin{abstract}
The effect of decoherence on a qubit-qutrit system under the influence of
global and multilocal decoherence in non-inertial frames is investigated. It
is shown that the entanglement sudden death (ESD) can be avoided in
non-inertial frames in the presence of dephasing and bit-trit phase flip
channels for entire range of decoherence. However, ESD behaviour is seen for
higher level of decoherence in case of phase flip and bit-trit flip
channels. Irrespective of the entanglement degradation caused by the Unruh
effect, no ESD occurs for the dephasing environment. Furthermore, flipping
channels have approximately similar effect on the entanglement of the hybrid
system.\newline
\end{abstract}

\maketitle

\address{Department of Physics Quaid-i-Azam University \\
Islamabad 45320, Pakistan}

\date{\today}

\newline

\vspace*{1.0cm}

\vspace*{1.0cm}

%\date{\today}

%\newpage

\section{Introduction}

Quantum entanglement plays an important role in various quantum information
tasks, such as quantum teleportation [1] and quantum computation [2].
However entangled states are always coupled with the environment due to the
unavoidable decoherence. As a result of the system-environment interaction,
the pure entangled states become mixed states and therefore become quite
useless for quantum information tasks [3]. Multi-partite entangled states of
qubits and qutrits are a central resource of quantum information science.
They can be used in constructing different protocols, for example, key
distribution and quantum computation [4, 5]. During recent years, two
important phenomenon, entanglement sudden death (ESD) and entanglement
sudden birth (ESB) have been investigated for bipartite and multipartite
states [6-13]. In this context, Yu and Eberly [14, 15] have investigated the
time evolution of entanglement of a bipartite qubit system undergoing
various modes of decoherence. Peres-Horodecki [16, 17] have studied
entanglement of qubit-qubit and qubit-qutrit states and established
separability criterion. According to this criterion, the partial transpose
of a separable density matrix must have non-negative eigenvalues, where the
partial transpose is taken over the smaller subsystem for qubit-qutrit case.
Ann et al. [18] have studied the ESD behaviour of a qubit-qutrit system
under the influence of dephasing noise.

Relativistic quantum information combines the tools from general relativity,
quantum field theory and quantum information theory. It is a rather new and
fast-growing field. It has attracted much attention during recent years to
study Unruh or Hawking effect on the entanglement shared between inertial
and non-inertial observers [19--30]. However, most of the investigations in
non-inertial frames are related to the study of the quantum information is
an isolated system. The decoherence [31], which appears when a system
interacts with its environment in a irreversible way, can be viewed as the
transfer of information from the system to its environment. It plays a
fundamental role in the description of the quantum-to-classical transition
[32] and has been successfully applied in the cavity QED [33] and ion trap
experiments [34]. Recently, implementation of decoherence in non-inertial
frames have been investigated for bipartite and tripartite systems [35, 36].

In this paper, the effect of decoherence on a qubit-qutrit system in
non-inertial frames by considering different noise models, such as, phase
flip, dephasing, bit-trit flip and bit-trit phase flip channels,
parameterized by decoherence parameter $p$ such that $p\in \lbrack 0,1]$.
The lower and upper limits of the decoherence parameter correspond to the
fully coherent and fully decohered system, respectively. Different couplings
of the system and the environment where the system is influenced by global
and multi-local noises are considered. It is shown that the ESD can be
avoided in non-inertial frames in the presence of dephasing and bit-trit
phase flip environments. However, no ESD occurs for the dephasing
environment.

\section{Qubit-qutrit system in non-inertial frames}

Let a composite system of qubit $A$ and qutrit $B$ is coupled to a noisy
environment both collectively and globally. Multi-local coupling describes
the situation when the qubit and qutrit are independently influenced by
their individual noisy environments. Whereas, the global decoherence
corresponds to the situation when it is influenced by both collective and
multilocal noises at the same time. The term collective coupling means when
both the qubit and qutrit are influenced by the same noise. In this study, a
particular initial state of the qubit-qutrit system is considered.

The density matrix can be computed by taking into account that Rob is
constrained to region $I$ of Rindler space-time, that requires to rewrite
Rob's mode in terms of Rindler modes and then to trace over the unobservable
Rindler's region $IV$. Therefore, the state shared between the two parties
is an Unruh-Rindler hybrid (qubit-qutrit system) entangled state as given by
[37]%
\begin{equation}
\rho _{AR}=\frac{1}{2}\left. \left(
\begin{array}{c}
\cos ^{2}r(|01\rangle _{AR}\left\langle 01\right\vert +|01\rangle
_{AR}\left\langle 10\right\vert +|10\rangle _{AR}\left\langle 01\right\vert
+|10\rangle _{AR}\left\langle 10\right\vert ) \\
+\sin ^{2}r(|02\rangle _{AR}\left\langle 02\right\vert +|12\rangle
_{AR}\left\langle 12\right\vert )%
\end{array}%
\right) .\right.
\end{equation}%
Alice has a qubit and Rob would have a qutrit, since for Rob's mode he could
have three different possible orthogonal states: particle spin-up, particle
spin-down and particle pair and the Minkowski-Rindler modes, the subscripts $%
A$ and $R$ correspond to Alice and Rob respectively. The notation $0$, $1$,
and $2$ correspond to spin up $\uparrow ,$ spin down $\downarrow $ and
spin-up/down $\uparrow \downarrow $ states respectively.

Let us assume that Alice remain stationary while Rob moves with uniform
acceleration. It is important to note that Rob, when he is accelerated with
respect to an inertial observer of the Dirac vacuum would observe a thermal
distribution of fermionic spin $1/2$ particles when he observes the
Minkowski vacuum due to Unruh effect [37]. The above state is obtained after
taking the trace over unobserved region $IV$ [38]. The evolution of a state
of a quantum system in a noisy environment can be described by the
super-operator in the Kraus operator representation as [39]

\begin{equation}
\rho _{f}=\sum_{k}E_{k}\rho _{i}E_{k}^{\dag },  \label{E5}
\end{equation}%
where the Kraus operators $E_{i}$ satisfy the following completeness relation

\begin{equation}
\sum_{k}E_{k}^{\dag }E_{k}=I.  \label{5}
\end{equation}%
The single qubit Kraus operators for phase flip, dephasing, bit flip and bit
phase flip channels are respectively given in table 1. Whereas the single
qutrit Kraus operators for the phase flip channel are given by

\begin{equation}
E_{0}=\left(
\begin{array}{ccc}
1 & 0 & 0 \\
0 & \sqrt{1-p} & 0 \\
0 & 0 & \sqrt{1-p}%
\end{array}%
\right) ,\ \ E_{1}=\left(
\begin{array}{ccc}
0 & \sqrt{p} & 0 \\
0 & 0 & 0 \\
0 & 0 & 0%
\end{array}%
\right) ,\ \ E_{2}=\left(
\begin{array}{ccc}
0 & 0 & \sqrt{p} \\
0 & 0 & 0 \\
0 & 0 & 0%
\end{array}%
\right)  \label{E7}
\end{equation}%
and the single qutrit Kraus operators for dephasing channel are given as%
\begin{equation}
E_{0}=\left(
\begin{array}{ccc}
1 & 0 & 0 \\
0 & \sqrt{1-p} & 0 \\
0 & 0 & \sqrt{1-p}%
\end{array}%
\right) ,\ \ E_{1}=\left(
\begin{array}{ccc}
1 & 0 & 0 \\
0 & \sqrt{p} & 0 \\
0 & 0 & 0%
\end{array}%
\right) ,\ \ E_{2}=\left(
\begin{array}{ccc}
1 & 0 & 0 \\
0 & 0 & 0 \\
0 & 0 & \sqrt{p}%
\end{array}%
\right)
\end{equation}%
Whereas the single qutrit Kraus operators for trit flip channel are given by

\begin{equation}
E_{0}=\sqrt{1-\frac{2p}{3}}\left(
\begin{array}{ccc}
1 & 0 & 0 \\
0 & 1 & 0 \\
0 & 0 & 1%
\end{array}%
\right) ,\ \ E_{1}=\sqrt{\frac{p}{3}}\left(
\begin{array}{ccc}
0 & 0 & 1 \\
1 & 0 & 0 \\
0 & 1 & 0%
\end{array}%
\right) ,\ \ E_{2}=\sqrt{\frac{p}{3}}\left(
\begin{array}{ccc}
0 & 1 & 0 \\
0 & 0 & 1 \\
1 & 0 & 0%
\end{array}%
\right)  \label{7}
\end{equation}%
and the single qutrit Kraus operators for the trit phase flip channel are
given by

\begin{eqnarray}
E_{0} &=&\sqrt{1-\frac{2p}{3}}\left(
\begin{array}{ccc}
1 & 0 & 0 \\
0 & 1 & 0 \\
0 & 0 & 1%
\end{array}%
\right) ,\ \ E_{1}=\sqrt{\frac{p}{3}}\left(
\begin{array}{ccc}
0 & 0 & e^{\frac{2\pi i}{3}} \\
1 & 0 & 0 \\
0 & e^{\frac{-2\pi i}{3}} & 0%
\end{array}%
\right) ,  \notag \\
E_{2} &=&\sqrt{\frac{p}{3}}\left(
\begin{array}{ccc}
0 & e^{\frac{-2\pi i}{3}} & 0 \\
0 & 0 & e^{\frac{2\pi i}{3}} \\
1 & 0 & 0%
\end{array}%
\right) ,\ \ E_{3}=\sqrt{\frac{p}{3}}\left(
\begin{array}{ccc}
0 & e^{\frac{2\pi i}{3}} & 0 \\
0 & 0 & e^{\frac{-2\pi i}{3}} \\
1 & 0 & 0%
\end{array}%
\right)
\end{eqnarray}%
The evolution of the initial density matrix of the composite system when it
is influenced by local and multi-local environments is given in Kraus
operator form as
\begin{equation}
\rho _{f}=\sum\limits_{i,j,k}(E_{j}^{B}E_{k}^{A})\rho
_{AR}(E_{j}^{B}E_{k}^{A})^{\dagger }
\end{equation}%
and the evolution of the system when it is influenced by global environment
is given in Kraus operator representation as
\begin{equation}
\rho _{f}=\sum\limits_{i,j,k}(E_{i}^{AB}E_{j}^{B}E_{k}^{A})\rho
_{AR}(E_{i}^{AB}E_{j}^{B}E_{k}^{A})^{\dagger }
\end{equation}%
where $E_{k}^{A}=E_{m}^{A}\otimes I_{3},$ $I_{2}\otimes E_{j}^{B}$ are the
Kraus operators of the multilocal coupling of qubit and qutrit individually
and $E_{i}^{AB}=E_{m}^{A}\otimes E_{n}^{A}$ are the Kraus operators of the
collective coupling of the qutrit system. Using equations (4-9) along with
the initial density matrix of as given in equation (1) and taking the
partial transpose over the smaller subsystem (qubit), the eigenvalues of the
final density matrix can be easily found. Let the decoherence parameters for
local and global noise of the qubit and qutrit be $p_{1}$, $p_{2}$ and $p$
respectively. The entanglement for all mixed states $\rho _{AB}$ of a
qubit-qutrit system is well quantified by the negativity [40]%
\begin{equation}
N(\rho _{AB})=\max \{0,\sum\limits_{k}\left\vert \lambda
_{k}^{T_{A}(-)}\right\vert \})
\end{equation}%
where $\lambda _{k}^{T_{A}(-)}$\ represents the negative eigenvalues of the
partial transpose of the density matrix $\rho _{AB}$ with respect to the
smaller subsystem.

\section{Results and Discussions}

The only possible negative eigenvalues of the partial transpose matrix when
the system is influenced by the multi-local and global noises of the phase
flip channel are given by
\begin{eqnarray}
\lambda _{\text{ml}} &=&-\frac{1}{2}\sqrt{(\text{$p_{1}$}-1)^{2}(\text{$%
p_{2} $}-1)^{2}\cos ^{4}(r)}  \notag \\
\lambda _{\text{g}} &=&\left(
\begin{array}{c}
-\frac{1}{2}\sqrt{(p-1)^{2}(\text{$p_{1}$}-1)^{4}(\text{$p_{2}$}-1)^{2}\cos
^{4}(r)}%
\end{array}%
\right)
\end{eqnarray}%
where the subscripts ml and g represent the multi-local and global noise
respectively. The negativity can be calculated using equation (10) for all
the cases under consideration. The only possible negative eigenvalues of the
partial transpose matrix when the system is influenced by the multi-local
and global noises of the dephasing channel are given by%
\begin{equation}
\begin{tabular}{l}
$\lambda _{\text{ml}}=-\frac{1}{2}\sqrt{(1-\text{$p_{1}$})\left( 2\sqrt{(1-%
\text{$p_{2}$})\text{$p_{2}$}}+1\right) \cos ^{4}(r)}$ \\
$\lambda _{\text{g}}=%
\begin{array}{c}
-\frac{1}{2}\sqrt{\left( 2\sqrt{(1-p)p}+1\right) (\text{$p_{1}$}%
-1)^{2}\left( 2\sqrt{(1-\text{$p_{2}$})\text{$p_{2}$}}+1\right) \cos ^{4}(r)}%
\end{array}%
$%
\end{tabular}%
\end{equation}

The relations for other channels are too lengthy and instead of presenting
in the text, I have potted them for analysis. In order to investigate the
effect of decoherence on the qubit-qutrit system in non-inertial frames, the
negativity is plotted as a function of decoherence parameter, $p$ in figure
1 (a) for Rob's acceleration $r$\ $=\pi /6$ (b) $r$\ $=\pi /4$ and as a
function of Rob's acceleration, $r$ (c) for $p=0.3$ and (d) $p=0.7$ for
multi-local noise. Here the decoherence parameter, $p$ corresponds to $%
p_{1}=p_{2}=p.$ It is seen that for multi-local coupling, no ESD is seen
even at $r$\ $=\pi /4$ for $p<1$. In figure2, negativity is plotted as a
function of decoherence parameter, $p$ in figure 2 (a) for Rob's
acceleration $r$\ $=\pi /6$ (b) $r$\ $=\pi /4$ and as a function of Rob's
acceleration, $r$ (c) for $p=0.3$ and (d) $p=0.7$ for global noise of
different channels. It is seen that in case of global noise, ESD behaviour
is seen for $p>0.6$ for phase flip and bit-trit flip channels. It is also
seen that maximal entanglement degradation occurs under global noise. It is
also seen that the entanglement is degraded heavily as we increase the value
of Rob's acceleration from $r$\ $=\pi /10$ to $r$\ $=\pi /4$ (infinite
acceleration limit). Furthermore, a similar behaviour of flipping channels
is seen towards entanglement degradation.

In figure 3, three dimensional graphs for negativity are plotted as a
function of Rob's acceleration, $r$\ and decoherence parameter, $p$
influenced by global noise of all the channels under consideration. It is
seen that for higher values of decoherence parameters, entanglement sudden
death occurs in case of flipping channels for $p>0.5$. Further more, it is
seen that no ESD occurs for any acceleration of Rob for the entire range of
decoherence parameters in case of dephasing channel.

\section{Conclusions}

The effect of decoherence on a qubit-qutrit system under the influence of
global and multilocal decoherence in non-inertial frames is investigated by
considering phase flip, dephasing, bit-trit flip and bit-trit phase flip
channels. It is shown that the entanglement sudden death can be avoided in
non-inertial frames in the presence of dephasing and bit-trit phase flip
channels for entire range of decoherence. ESD behaviour is seen for higher
level of decoherence in case of phase flip and bit-trit flip channels.
However, no ESD occurs for the dephasing environment. Furthermore, flipping
channels have a similar effect on the entanglement of the hybrid system.

{\huge Figures captions}\newline
\textbf{Figure 1}. (Color online). The negativity is plotted as a function
of decoherence parameter, $p$ in figure 1 (a) for Rob's acceleration $r$\ $%
=\pi /6$ (b) $r$\ $=\pi /4$ and as a function of Rob's acceleration, $r$ (c)
for $p=0.3$ and (d) $p=0.7$ for multi-local noise.\newline
\textbf{Figure 2}. (Color online). The negativity is plotted as a function
of decoherence parameter, $p$ in figure 2 (a) for Rob's acceleration $r$\ $%
=\pi /6$ (b) $r$\ $=\pi /4$ and as a function of Rob's acceleration, $r$ (c)
for $p=0.3$ and (d) $p=0.7$ for global noise of different channels.\newline
\textbf{Figure 3}. (Color online). The negativity are plotted as a function
of Rob's acceleration, $r$\ and decoherence parameter, $p$ influenced by
global noise of all the channels under consideration.\newline
{\Huge Table Caption}\newline
\textbf{Table 1}. Single qubit Kraus operators for phase flip, dephasing,
bit-trit flip and bit-trit phase flip channels where $p$ represents the
decoherence parameter.\newline
\newpage

\begin{figure}[tbp]
\begin{center}
\vspace{-2cm} \includegraphics[scale=0.6]{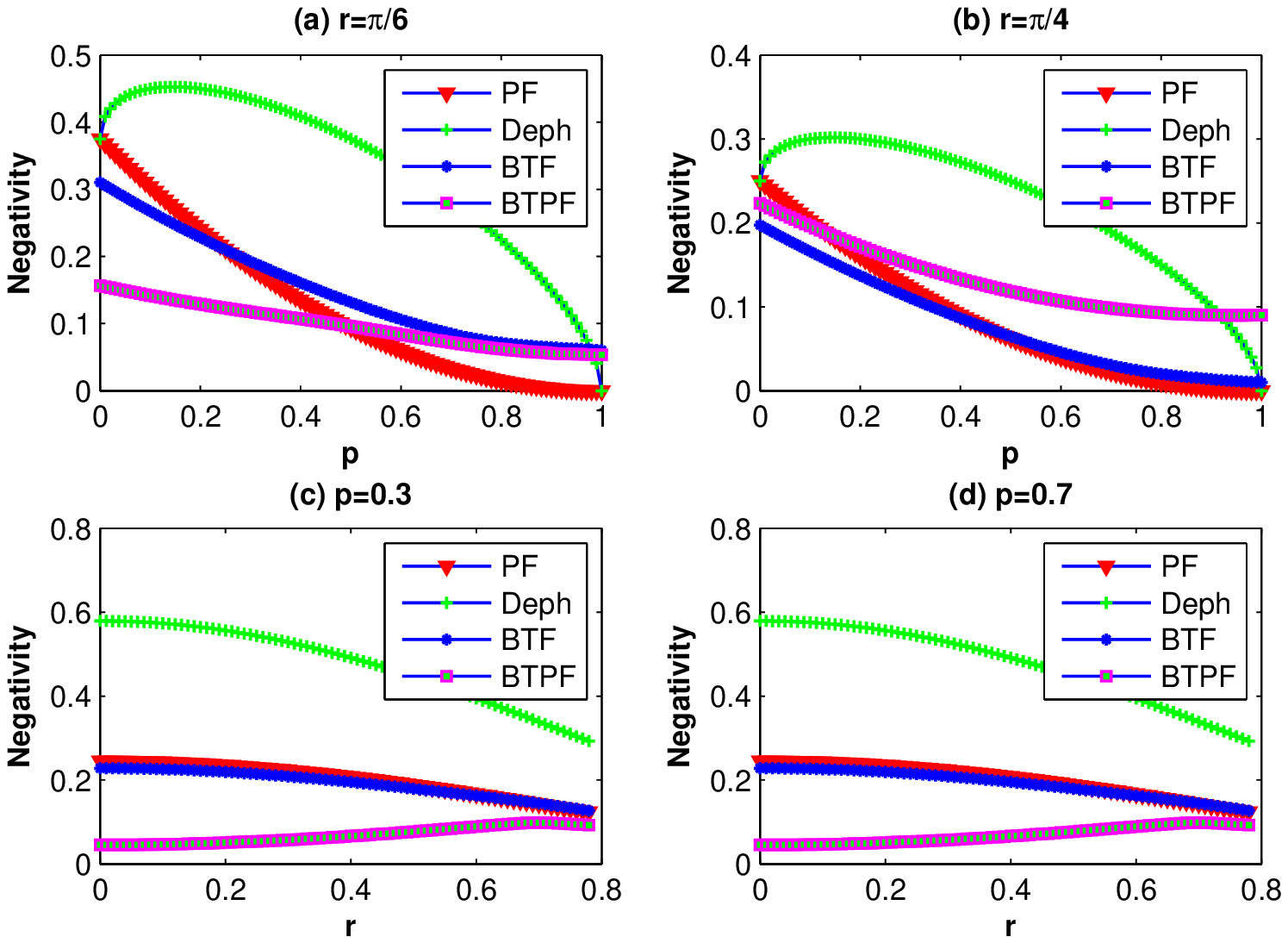} \\[0pt]
\end{center}
\caption{(Color online). The negativity is plotted as a function of
decoherence parameter, $p$ in figure 1 (a) for Rob's acceleration $r$\ $=%
\protect\pi /6$ (b) $r$\ $=\protect\pi /4$ and as a function of Rob's
acceleration, $r$ (c) for $p=0.3$ and (d) $p=0.7$ for multi-local noise.}
\end{figure}
\newpage

\begin{figure}[tbp]
\begin{center}
\vspace{-2cm} \includegraphics[scale=0.6]{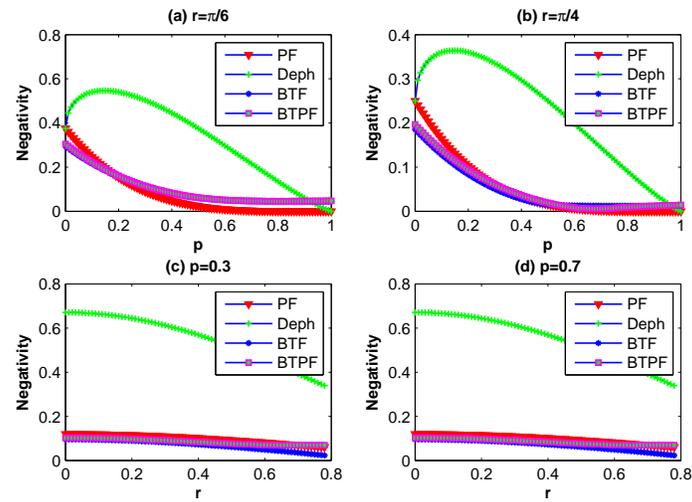} \\[0pt]
\end{center}
\caption{(Color online). The negativity is plotted as a function of
decoherence parameter, $p$ in figure 2 (a) for Rob's acceleration $r$\ $=%
\protect\pi /6$ (b) $r$\ $=\protect\pi /4$ and as a function of Rob's
acceleration, $r$ (c) for $p=0.3$ and (d) $p=0.7$ for global noise of
different channels.}
\end{figure}
\newpage

\begin{figure}[tbp]
\begin{center}
\vspace{-2cm} \includegraphics[scale=0.6]{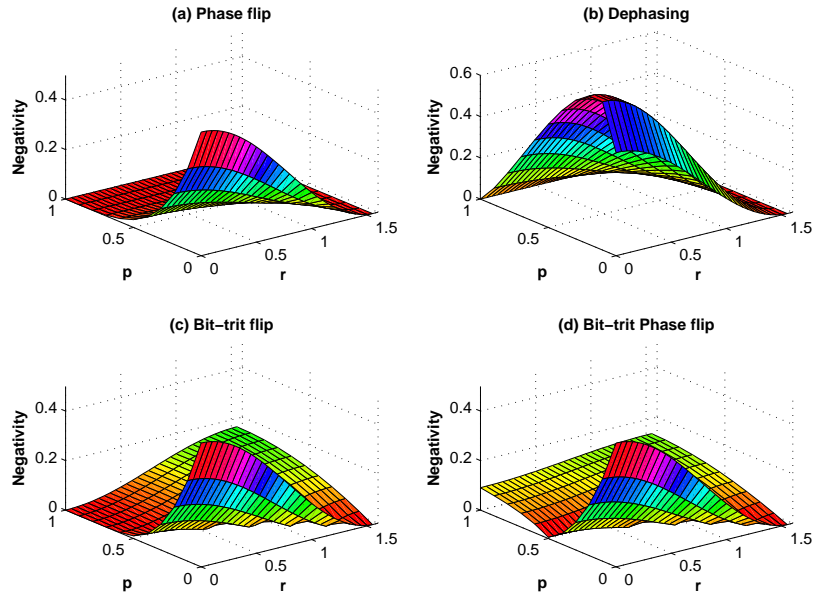} \\[0pt]
\end{center}
\caption{(Color online). The negativity are plotted as a function of Rob's
acceleration, $r$\ and decoherence parameter, $p$ influenced by global noise
of all the channels under consideration.}
\end{figure}

\newpage

\begin{table}[tbh]
\caption{Single qubit Kraus operators for phase flip, dephasing, bit-trit
flip and bit-trit phase flip channels where $p$ represents the decoherence
parameter.}
\label{di-fit}$%
\begin{tabular}{|l|l|}
\hline
&  \\
Phase flip$\text{ channel}$ & $E_{0}=\sqrt{1-\frac{p}{2}I},\quad E_{1}=\sqrt{%
\frac{p}{2}}\left[
\begin{array}{cc}
1 & 0 \\
0 & -1%
\end{array}%
\right] $ \\
&  \\ \hline
&  \\
Dephasing$\text{ channel}$ & $E_{0}=\left[
\begin{array}{cc}
1 & 0 \\
0 & \sqrt{1-p}%
\end{array}%
\right] ,$ $E_{1}=\left[
\begin{array}{cc}
0 &  \\
0 & \sqrt{p}%
\end{array}%
\right] $ \\
&  \\ \hline
&  \\
&  \\ \hline
&  \\
$\text{Bit flip channel}$ & $E_{0}=\sqrt{1-\frac{p}{2}}I,\quad E_{1}=\sqrt{%
\frac{p}{2}}\left[
\begin{array}{cc}
0 & 1 \\
1 & 0%
\end{array}%
\right] $ \\
&  \\ \hline
&  \\
$\text{Bit-phase flip channel}$ & $E_{0}=\sqrt{1-\frac{p}{2}}I,\quad E_{1}=%
\sqrt{\frac{p}{2}}\left[
\begin{array}{cc}
0 & -i \\
i & 0%
\end{array}%
\right] $ \\
&  \\ \hline
\end{tabular}%
$%
\end{table}


\begin{thebibliography}{99}
\bibitem{1} C. H. Bennett, G. Brassard, C. Crepeau, R. Jozsa, A. Peres, and
W. Wootters, Phys. Rev. Lett. 70, 1895 (1993).

\bibitem{2} H. J. Briegel and R. Raussendorf Phys. Rev. Lett. 86, 910 (2001).

\bibitem{3} M. Siomau and S. Fritzsche, quant-ph/1101.3275 (2011).

\bibitem{KD} Ekert, A.: Phys. Rev. Lett. \textbf{67,} 661-663 (1991).

\bibitem{QC} Grover, L.K.: Phys. Rev. Lett. \textbf{79,} 325-328 (1997).

\bibitem{ESD1} Yonac, M., et al.: J. Phys. B \textbf{39,} S621-S625 (2006).

\bibitem{ESD2} Jakobczyk, L., Jamroz, A.:Phys. Lett. A \textbf{333,} 35-45
(2004).

\bibitem{ESD3} Ann, K., Jaeger, G.: Phys. Rev. A \textbf{76,} 044101 (2007).

\bibitem{ESD4} Jaeger, G., Ann, K.: J. Mod. Opt. \textbf{54,} 2327-2338
(2007).

\bibitem{10} A. R. P. Rau, M. Ali and G. Albert \ Europhysics Letter 82,
4002 (2009).

\bibitem{11} Z.-G. Li. et al. Phys. Rev. A 79, 024303 (2009).

\bibitem{ESB1} Ficek, Z., Tanas, R.: Phys. Rev. A \textbf{77,} 054301 (2008).

\bibitem{ESB2} Lopez, C.E., Romero, G., Lastra, F., Solano, E., Retamal,
J.C.: Phys. Rev. Lett. \textbf{101,} 080503 (2008).

\bibitem{HG} Yu, T., Eberly, J.H.: Phys. Rev. B \textbf{68,} 165322 (2003).

\bibitem{HB} Yu, T., Eberly, J.H.: Phys. Rev. Lett. \textbf{97,} 140403
(2007).

\bibitem{PR} Peres, A.: Phys. Rev. Lett. \textbf{77,} 1413-1415 (1996).

\bibitem{HD} Horodecki, M., Horodecki, P., Horodecki, R.: Phys. Lett. A
\textbf{223,} 1-8 (1996).

\bibitem{Ann} Ann, K., Jaeger, G.: Phys. Lett. A \textbf{372,} 579-583
(2008).

\bibitem{AB} Alsing, P.M., Milburn, G.J.: Phys. Rev. Lett. \textbf{91},
180404 (2003).

\bibitem{AC} Fuentes-Schuller, I., Mann, R.B.: Phys. Rev. Lett. \textbf{95},
120404 (2005).

\bibitem{AD} Alsing, P.M., Fuentes-Schuller, I., Mann, R.B., Tessier, T.E.:
Phys. Rev. A \textbf{74}, 032326 (2006).

\bibitem{AG} Ralph, T.C., Milburn, G.J., Downes, T.: Phys. Rev. A \textbf{79}%
, 022121 (2009).

\bibitem{AH} Doukas, J., Hollenberg, L.C.L.: Phys. Rev. A \textbf{79},
052109 (2009).

\bibitem{AJ} Moradi, S.: Phys. Rev. A \textbf{79}, 064301 (2009).

\bibitem{AK} Martn-Martnez, E., Len, J.: Phys. Rev. A \textbf{80}, 042318
(2009).

\bibitem{AL} Wang, J., Deng, J., Jing, J.: Phys. Rev. A \textbf{81}, 052120
(2010).

\bibitem{AW} Pan, Q., Jing, J.: Phys. Rev. A \textbf{77}, 024302 (2008).

\bibitem{AR} Wang, J., Pan, Q., Jing, J.: Phys. Lett. B \textbf{692}, 202
(2010).

\bibitem{nif} Montero, M., Martin-Martinez, E.: arXiv:quant-ph/1011.6540
(2011).

\bibitem{nif2} David, E., et al.: Phys. Rev. A \textbf{82}, 042332 (2010).

\bibitem{29} W. H. Zurek, Rev. Mod. Phys. 75, 715 (2003).

\bibitem{30} M. A. Schlosshauer, Decoherence and the Quantum-To- Classical
Transition (Springer), (2007).

\bibitem{31} M. Brune, E. Hagley, J. Dreyer, X. Maitre, A. Maali, C.
Wunderlich, J. M. Raimond, and S. Haroche, Phys. Rev. Lett. 77, 4887 (1996).

\bibitem{32} C. J. Myatt, B. E. King, Q. A. Turchette, C. A. Sackett, D.
Kielpinski,W.M. Itano, C.Monroe, and D. J.Wineland, Nature 403, 269 (2000).

\bibitem{EisJ2} Wang, J., Jing, J.: Phys. Rev. A \textbf{82,} 032324 (2010).

\bibitem{Sal} Ramzan M and Khan, M.K. Quant. Inf. Proc. DOI
10.1007/s11128-011-0257-7 (2011).

\bibitem{UE} Unruh, W.G.: Notes on black-hole evaporation. Phys. Rev. D
\textbf{14}, 870 (1976).

\bibitem{SON} Leon, J., Martin-Martinez, E.: Phys. Rev. A \textbf{80},
012314 (2009).

\bibitem{NC} Nielson, M.A., Chuang, I.L.: Quantum Computation and Quantum
Information (Cambridge: Cambridge University Press), (2000).

\bibitem{VD} Vidal, G., Werner, R.F.: Phys. Rev. A \textbf{65}, 032314
(2002).\newpage
\end{thebibliography}
\end{document}